\documentstyle[12pt]{article}
\begin{document}

\title{Gravitomagnetic effect and spin-torsion coupling}
\author{A. A. Sousa* \\
Departamento de Matem\'{a}tica\\
Instituto de Ci\^{e}ncias e Letras do M\'{e}dio Araguaia \\
Universidade Federal de Mato Grosso\\
78698-000 Pontal do Araguaia, MT, Brazil\\
and\\
J. W. Maluf$^{\S }$\\
Instituto de F\'{\i}sica, Universidade de Bras\'{\i}lia \\
70.919-970 Bras\'{\i}lia, DF, Brazil\\
}
\maketitle

\begin{abstract}
We study the gravitomagnetic effect in the context of absolute parallelism
with the use of a modified geodesic equation via a free parameter $b$. We
calculate the time difference in two atomic clocks orbiting the Earth in
opposite directions and find a small correction due the coupling between the
torsion of space time and the internal structure of atomic clocks measured
by the free parameter.

PACS NUMBERS: 04.80.Cc, 04.90.+e

(*) e-mail: adellane@cpd.ufmt.br

($\S $) e-mail: wadih@fis.unb.br
\end{abstract}

\section{Introduction}

In attempting to seek experimental confirmation of the gravitomagnetic
effect suggested by Mashhoon {\it et al}. $\cite{1}$, the called Gravity
Probe C(lock) experiment was proposed $\cite{2}$. The experimental
confirmation is considered difficult due several perturbations of planetary
origin that can hide the gravitomagnetic effect $\cite{3}$. In the
experiment a clock is sent in a direct equatorial and circular orbit, and
another clock in a retrograde orbit, both clocks considered without internal
structure. The time difference marked by the clocks is expected to be $(4\pi
a)/{c}\sim 2,327\times 10^{-7}s$. This difference is considered for an
exterior observer with $r\gg $ $(2GM)/{c^{2}}$, where $a=J/Mc$, $J$ is the
Earth angular momentum, $M$ is the mass of the Earth, $G$ is the
Gravitational constant and $c$ is the speed of light. This effect is
interpreted as the dragging of a inertial frame due to the Earth's rotation.

It is well-known that photon's trajectories are of fundamental importance
for astronomy in several observed waves lengths. Photons move in space-time
according to the geodesic equation, where the Christoffel connection plays a
fundamental role. This geodesic equation embodies Einstein's equivalence
principle. The trajectories of the photons in the space-time are used in the
explanation of the following classical tests of general relativity: the test
of the redshift, the light rays deflection and the time delay of radar
signals around planets, known as Shapiro effect. In these tests, the photons
are treated as light rays, that is, particles without spin. Investigations
carried out with the help of the parametrized post-Newtonian formalism (PPN)
suggest several observations to prove these tests. They produced results
that prove the predictions of general relativity with high precision $\cite
{4}$. However, although the agreement favours general relativity, it does
not mean that some corrections to the theory cannot be implemented,
corrections that yield results which agree with the experimental error
limits. Thus it is possible that the general relativity theory be a correct
gravitational theory within certain limits.

Wanas and Kahil $\cite{5}$ and Wanas {\it et al.} $\cite{6}$ proposed to
explain the discrepancy between the thermal neutrons interference experiment
and the theoretical prediction, by means of Bazanski's formalism $\cite{7}$,
through the ``quantization'' of the path followed by the particles with
spin. They used a modified geodesic equation to include Einstein's absolute
parallelism using a nonsymmetrical connection. They applied this equation to
the weak field limit and found that the Newtonian gravitational potential is
modified for a factor $(1-b)$, where $b$ establishes the coupling between
the torsion field and the intrinsic spin. For particles with spin, they
postulate that $b=(n/2)\alpha \gamma $, where $\alpha $ is the fine
structure constant and $\gamma $ is a parameter to match with the
experience. For $n=0,1,2,3...$ the particles assume spin $zero,1/2,1,3/2$
etc. For macroscopic bodies (without spin), $n=0$ and $b=0$. This
interaction would take place through the coupling of the spin particle with
the space-time torsion. However, new experiments would need to be
accomplished to test such quantization of the path.

In this article we propose a new test to verify the gravitomagnetic effect
and Wanas' conclusions by considering a covariant derivative definition $%
D_{\mu }e_{av}=0$ in the absolute parallelism framework, that yields a class
of geodesic equations, and taking into account the identity $\omega _{\mu
ab}=$ $^{0}\omega _{\mu ab}+K_{\mu ab}$, where $\omega _{\mu ab}$ is an
arbitrary affine connection, $e_{a\mu }$ is tetrad field with Lorentz
indices $a,b,...\,$, $^{0}\omega _{\mu ab}$ is the Levi-Civita connection
and $K_{\mu ab}$ is the contortion tensor. We impose the time gauge
condition $\cite{8}$ for the tetrad field by fixing $e_{(k)}\,^{0}=0$ and $%
e^{(0)}\,_{k}=0$. We find the same geodesic equation obtained by Wanas by
assuming that for particles with nonzero spin the violation of the
equivalence principle is negligible, and therefore the coupling with the
torsion is very small. The latter takes place by means of an empirical
parameter $b$, that characterizes the coupling between torsion and the spin
of particles. When applying this new equation to the Kerr metric, for a
circular and equatorial orbit, we find that the period difference measured
by the clocks is about $[(4\pi a)/c][(1-2b)/(1-b)]$. 
We also conclude that the orbital period is given by $T_{o}=\frac{2\pi }{%
\omega _{0}(1-b)^{1/2}}$, which is larger than the expected value and
indicates that the Newtonian gravitational potential is modified by means of
a factor $(1-b)$, namely, $\phi =-\frac{GM}{r}(1-b)$. Therefore there is a
modification of order $(1-b)^{1/2}$ in the Keplerian period. The reason for
this is that the potential on the clock must be smaller than the usual
Newtonian potential by a factor $(1-b)$. The clock would be under the action
of a smaller potential, with a smaller acceleration, registering a longer
time to complete an orbit. When considering clocks without internal
structure, as done previously, we make $b=0$. Due to the fact that
measurements of time differences of such low order
require the use atomics
clocks, as H maser (maser of Hydrogen) and Cesium $133$, we suggest a
coupling of the internal structure of these clocks with the space-time
torsion. The frequency of the Cs atomic clock is $\nu _{0}\approx 9.2$ GHz
and corresponds to the $\lbrack F=4,\;\; m_{F}=0\rbrack
\rightarrow \lbrack F=3,\;\; m_{F}=0\rbrack$ hyperfine
transition in the $^{133}$Cs ground state. As for the H maser there
corresponds the $\lbrack F=1,\;\; m_{F}=0\rbrack
\rightarrow \lbrack F=0,\;\; m_{F}=0\rbrack$ transition with frequency
$ \nu _{0}\approx 1.4$ GHz. The violation of local position invariance
incorporated in Einstein's equivalence principle can be used to quantify
the dimensionless parameter $\beta $ (positive or negative) that measures the
discrepancy between the observed and predicted redshift $\Delta \nu $ of
spectral lines of atomic clocks (see, for example, $\cite{4}$ and references
therein). The parameter $\beta $ depends on the nature of the measured clock.
The parameter $b$ can be determined in the same manner
by means of  the expression
$\Delta \nu /\nu _{0}=\left[ 1+(1-b)\phi _{N}/c^{2}\right] $, where $\phi
_{N} $ $=-GM/r$ is
the Newtonian gravitational potential. For two identical 
$^{133}$Cs clocks the result is $\left| b\right| $ $<1.5\times
10^{-2}$. Consequently this value represent a $1.52\%$ difference
with respect to Mashhoon's prediction.

We have considered the weak field approximation of modified geodesic
equations that satisfy the Newtonian limit in an arbitrary teleparallel
theory. The latter theory is defined to be quadratic in the torsion tensor
with free parameters $c_{1}$, $c_{2}$ and $c_{3}$ $\cite{9}$, $\cite{10}$.
The condition of Legendre transform for a well defined Hamiltonian
formulation is given by $c_{1}+c_{2}=0$. In the present context we found
that $c_{1}=-\frac{2}{3}k\frac{1}{(1-b)}$, $c_{2}=\frac{2}{3}k\frac{1}{(1-b)}
$ and $c_{3}=-\frac{3}{2}k\frac{1}{(1-b)}$, where $k=\frac{c^{3}}{16\pi G}$,
indicating that if we consider particles without internal structure (spin),
then $b=0$, resulting in the teleparallel equivalent of general relativity.
Thus $b$ cannot be $1$, what it is agreement with the fact that the torsion
coupling with spin must be small, resulting in a small violation of the
principle of equivalence.

In section 2 we review the geodesic equation for particles in a
gravitational field with a symmetric connection. In section 3, we introduce
the tetrad field description of the Weitzenb\"{o}ck space-time. In section 4
we carry out the calculations of the gravitomagnetic effect with a
nonsymmetric connection via the free empirical parameter $b$, displaying the
difference with respect to general relativity. In the section 5 we
provide estimates of the empirical parameter $b$. 
In section 6, we introduce the relationship with the
teleparallel equivalent of general relativity in the weak field
approximation. In section 7 the conclusions are presented.

The notation is the following: space-time indices $\mu ,$ $\nu $, ...and $%
SO(3,1)$ Lorentz indices $a,b$,... run from $0$ to $3$. In the 3+1
decomposition Latin indices from the middle of the alphabet indicate space
indices according to $\mu =0,i$ and $a=(0),(i)$. The flat space-time metric
is fixed by $\eta _{(0)(0)}=-1$.


\section{The gravitomagnetic effect in the general relativity}

The exterior space-time of a system with mass $M$ and specific angular
momentum $a=J/M$ is described by the Kerr geometry. The Kerr metric is an
exact solution of the vacuum field equations of general relativity. Written
in Boyer-Lindquist coordinates $(t,r,\theta ,\varphi )$, the Kerr metric
reads

\begin{equation}
ds^{2}=-dt^{2}+\Sigma \left( \frac{1}{\Delta }dr^{2}+d\theta ^{2}\right)
+\left( r^{2}+a^{2}\right) \sin ^{2}\theta d\varphi ^{2}+2M\frac{r}{\Sigma }%
\left( dt-a\sin ^{2}\theta \right) ^{2},  \label{1}
\end{equation}
where we have $\Sigma =r^{2}+a^{2}\cos ^{2}\theta $ and $\Delta
=r^{2}-2Mr+a^{2}$, and in this section $G=c=1.$

We first calculate the time registered by a standard clock that follows a
geodesic in the Kerr geometry $\cite{1}$. We choose a circular and
equatorial \ orbit$.$ The geodesic equation results in

\begin{equation}
dt^{2}-2ad\varphi dt+\left( a^{2}-\frac{r^{3}}{M}\right) d\varphi ^{2}=0,
\end{equation}
whose solutions are

\begin{equation}
\frac{dt}{d\varphi }=a\pm \left( \frac{r^{3}}{M}\right) ^{1/2}=a\pm \frac{1}{%
\omega _{0}},  \label{35}
\end{equation}
where $\omega _{0}$ is the keplerian angular velocity.

We find, with the help of $\left( \ref{1}\right) $, the relation

\begin{equation}
\left( \frac{d\tau }{d\varphi }\right) ^{2}=\left( 1-\frac{2M}{r}\right)
\left( \frac{dt}{d\varphi }\right) ^{2}+4\frac{Ma}{r}\frac{dt}{d\varphi }%
-r^{2}-a^{2}\left( 1+2\frac{M}{r}\right) .
\end{equation}
Substituting $\left( \ref{35}\right) $ in the equation above we obtain, for
a closed orbit, and considering an observer at the infinity such that $r\gg $
$2M$,

\begin{equation}
\tau _{+}-\tau _{-}\approx 4\pi a=4\pi \frac{J}{M},
\end{equation}
where the signals $+$ and $-$ apply for a direct and retrograde orbit,
respectively.

Introducing the speed of light $c$,

\begin{eqnarray}
\tau _{+}-\tau _{-} &\approx &4\pi \frac{J}{Mc^{2}} \\
&\approx &2,327\times 10^{-7}s,  \nonumber
\end{eqnarray}
where we used $M\approx 6\times 10^{24}$ $kg$ and $J\approx 10^{34}$ $%
kg.m^{2}.s^{-1}$, the mass and angular momentum of the Earth, respectively.
We can now ask what would happen if the clock would follow a geodesic
different from the Riemannian one, for example, one due to a nonsymmetrical
connection.

\section{The Weitzenb\"{o}ck space-time}

We present now a brief summary of the space-time of the Riemann-Cartan type
that is endowed with a metric $g_{\mu \nu }$ and a connection $\Gamma _{\mu
\nu }^{\lambda }$.

The Riemann-Cartan space-time is characterized by $\cite{11}$

\begin{equation}
\nabla _{\lambda }g_{\mu \nu }=\partial _{\lambda }g_{\mu \nu }-\Gamma _{\mu
\lambda }^{\rho }g_{\rho \nu }-\Gamma _{\nu \lambda }^{\rho }g_{\mu \rho }=0.
\label{eq1}
\end{equation}
From this equation we obtain

\begin{equation}
\Gamma _{\mu \nu }^{\lambda } =\;^{0}\Gamma _{\mu \nu }^{\lambda }+K_{\, \mu
\nu }^{\lambda },  \label{eq2}
\end{equation}
where the first member on the right hand side is the Christoffel connection,

\begin{equation}
^{0}\Gamma _{\mu \nu }^{\lambda }=\frac{1}{2}g^{\lambda \rho }\left(
\partial _{\mu }g_{\nu \rho }+\partial _{\nu }g_{\mu \rho }-\partial _{\rho
}g_{\mu \nu }\right) ,  \label{eq3}
\end{equation}
and second term is the contortion tensor,

\begin{equation}
K^\lambda\,_{\mu\nu} =\frac{1}{2}\left( T^\lambda\,_{\mu\nu}
+T_\mu\,^\lambda\,_\nu -T_\nu\,^\lambda\,_\mu \right) .  \label{eq4}
\end{equation}
The torsion tensor is given by

\begin{equation}
T^\lambda\,_{\mu\nu} \left( \Gamma \right) =\Gamma _{\mu \nu }^{\lambda
}-\Gamma _{\nu \mu }^{\lambda },  \label{eq5}
\end{equation}
and the curvature tensor by

\begin{equation}
R^\mu\,_{\nu \alpha \beta }(\Gamma ) =\partial _{\alpha }\Gamma _{\beta \nu
}^{\mu }- \partial _{\beta }\Gamma _{\alpha \nu }^{\mu }+\Gamma _{\alpha
\sigma }^{\mu }\Gamma _{\beta \nu }^{\sigma } -\Gamma _{\beta \sigma }^{\mu
}\Gamma _{\alpha \nu }^{\sigma }.  \label{eq6}
\end{equation}

The Riemann-Cartan space-time is characterized by nonzero curvature and
torsion tensors. It leads to two geometrical models for the space-time. The
first is the Riemannian space-time, that is obtained by requiring the
vanishing of the torsion tensor. Therefore the space-time affine connection
reduces to the Christoffel connection. Another model is the Weitzenb\"{o}ck
space-time, that is obtained from Riemann-Cartan space-time by requiring the
curvature tensor to vanish,

\begin{equation}
R^\mu\,_{\nu \alpha \beta }(\Gamma )=0.  \label{eq7}
\end{equation}

The Weitzenb\"{o}ck space-time is endowed with the affine connection

\begin{equation}
\Gamma _{\mu \nu }^{\lambda }=e_{a}\,^{\lambda }\partial _{\mu }e^{a}\,_{\nu
}=-e^{a}\,_{\nu }\partial _{\mu }e_{a}\,^{\lambda }.  \label{eq20}
\end{equation}
where $e^{a}\,_{\mu }$ are orthonormal tetrads. The indices $a,b,c,...$ are
called local tetrads or indices of the $SO(3,1)$ group.

The affine connection $\left( \ref{eq20}\right) $ is not symmetrical with
respect to a change of the lower indices. Therefore the torsion tensor is
given by

\begin{eqnarray}
T^{\lambda }\,_{\mu \nu }\left( \Gamma \right) &=&\Gamma _{\mu \nu
}^{\lambda }-\Gamma _{\nu \mu }^{\lambda }=  \nonumber \\
&=&e_{a}\,^{\lambda }(\partial _{\mu }e^{a}\,_{\nu }-\partial _{\nu
}e^{a}\,_{\mu }).  \label{eq25}
\end{eqnarray}
From now on we will adopt the Weitzenb\"{o}ck space-time. In other words,
the space-time will be characterized by $R^{\rho }\,_{\sigma \mu \nu
}(\Gamma )=0$ and $T^{\lambda }\,_{\mu \nu }\left( \Gamma \right) \neq 0$.

\section{Gravitomagnetic effect with a nonsymmetrical connection}

In the Weitzenb\"{o}ck space-time the covariant derivative of the tetrad
field $e_{a\mu }$ vanish,

\begin{equation}
D_{\mu }e_{a\nu }=0,
\end{equation}
from what follows

\begin{equation}
e^{a\lambda }\partial _{\mu }e_{a\nu }=\,^{0}\Gamma _{\mu \nu }^{\lambda
}-e^{a\lambda }e^{b}\,_{\nu }\,^{0}\omega _{\mu ab},  \label{2}
\end{equation}
where $^{0}\omega _{\mu ab}$ is the Levi-Civita connection, which plays an
important role in the interaction of spin 1/2 matter fields with the
gravitational field. For an arbitrary connection $\omega _{\mu ab}$ there
exists the identity

\begin{equation}
\omega _{\mu ab}=\,^{0}\omega _{\mu ab}+K_{\mu ab},
\end{equation}
where $K_{\mu ab}$ is the contortion tensor. It follows that by fixing $%
\omega _{\mu ab}=0$, we obtain

\begin{equation}
^{0}\omega _{\mu ab}=-K_{\mu ab}=-\frac{1}{2}e_a\,^\lambda e_b\,^\nu \left(
T_{\lambda \mu \nu }+T_{\nu \lambda \mu }-T_{\mu \nu \lambda }\right) ,
\end{equation}
and

\begin{equation}
\,e^{a\lambda }e^b\,_\nu\,^{0}\omega _{\mu ab}= \frac{1}{2}g^{\lambda \rho
}\left( T_{\mu \nu \rho }+T_{\nu \mu \rho }\right) -\frac{1}{2}g^{\lambda
\rho }T_{\rho \mu \nu },
\end{equation}
which, except for the parameter $b$, allows us to rewrite equation $\left( 
\ref{2}\right) $ as

\begin{equation}
\Gamma _{\mu \nu }^{\lambda }=e^{a\lambda }\partial _{\mu }e_{a\nu
}=\,^{0}\Gamma _{\mu \nu }^{\lambda }-b\left[ \frac{1}{2}g^{\lambda \rho
}\left( T_{\mu \nu \rho }+T_{\nu \mu \rho }\right) +\frac{1}{2}g^{\lambda
\rho }T_{\rho \mu \nu }\right] .  \label{3}
\end{equation}
The empirical parameter $b$ has been introduced to account for observational
or experimental evidences. For $b=1$, the connection $\left( \ref{3}\right) $
reduces to Cartan's connection, describing the autoparallels (the
straightest curves in Riemann-Cartan space) $\cite{12}$. For $b=0$, we
recover the Christoffel connection together with the results of section $2$.

The fixation of $\omega _{\mu ab}=0$ seems to be important for a well
defined Hamiltonian formulation, and in order to have a correct time
evolution of the field quantities in the realm of the teleparallel
equivalent to the general relativity (TEGR) $\cite{10},\cite{13},\cite{14}$.

Assuming that we can test a new geodesic equation by substituting the
Christoffel connection by the nonsymmetrical connection $\left( \ref{3}%
\right) $, we can write

\begin{equation}
\frac{d^{2}x^{\lambda }}{d\tau ^{2}}+\,^{0}\Gamma _{\mu \nu }^{\lambda }%
\frac{dx^{\mu }}{d\tau }\frac{dx^{\nu }}{d\tau }-b\frac{g^{\lambda \rho }}{2}%
\left( T_{\mu \nu \rho }+T_{\nu \mu \rho }\right) \frac{dx^{\mu }}{d\tau }%
\frac{dx^{\nu }}{d\tau }=0.  \label{4}
\end{equation}
The equation above is identical to the one found by Wanas and Kahil $\cite{5}
$ using a variational principle in the context of Bazanski's formalism $\cite
{7}$ for the space-time of absolute parallelism.

By using equation $\left( \ref{4}\right) $ we are going to calculate the
time difference measured by a clock in direct orbit around the Earth and by
another one in retrograde orbit, and eventually we will compare the result
with that of section $2$. To this purpose, we are going to use the line
element

\begin{equation}
ds^{2}=-\frac{\Delta }{\rho ^{2}}\left( cdt- a\sin^{2}\theta d\varphi
\right) ^{2}+\frac{\sin ^{2}\theta }{\rho ^{2}}\left[ \left(
r^{2}+a^{2}\right) d\varphi -acdt\right] ^{2} +\frac{\rho ^{2}}{\Delta }%
dr^{2}+\rho ^{2}d\theta ^{2}.  \label{9}
\end{equation}
The metric in spherical coordinates is given by

\begin{equation}
g_{\mu \nu }=\left( 
\begin{array}{cccc}
-\frac{\Psi ^{2}}{\rho ^{2}} & 0 & 0 & -\frac{\chi \sin ^{2}\theta }{\rho
^{2}} \\ 
0 & \frac{\rho ^{2}}{\Delta } & 0 & 0 \\ 
0 & 0 & \rho ^{2} & 0 \\ 
-\frac{\chi \sin ^{2}\theta }{\rho ^{2}} & 0 & 0 & \frac{\Sigma ^{2}\sin
^{2}\theta }{\rho ^{2}}
\end{array}
\right) .  \label{8}
\end{equation}
We also have

\begin{equation}
g^{\mu \nu }=\left( 
\begin{array}{cccc}
\frac{-\rho ^{2}\Sigma ^{2}}{\Psi ^{2}\Sigma ^{2} +\chi ^{2}\sin ^{2}\theta }
& 0 & 0 & \frac{-\rho ^{2}\chi }{\Psi ^{2}\Sigma ^{2}+\chi ^{2}\sin
^{2}\theta } \\ 
0 & \frac{\Delta }{\rho ^{2}} & 0 & 0 \\ 
0 & 0 & \frac{1}{\rho ^{2}} & 0 \\ 
\frac{-\rho ^{2}\chi }{\Psi ^{2}\Sigma ^{2} +\chi ^{2}\sin ^{2}\theta } & 0
& 0 & \frac{\rho ^{2}\Psi ^{2}}{\left( \Psi ^{2}\Sigma ^{2}+ \chi ^{2}\sin
^{2}\theta \right) \sin ^{2}\theta }
\end{array}
\right) ,  \label{5}
\end{equation}
with the following definitions

\begin{eqnarray}
\Delta &=&r^{2}+a^{2}-2\frac{GM}{c^{2}}r, \\
\rho ^{2} &=&r^{2}+a^{2}\cos ^{2}\theta , \\
\Sigma ^{2} &=&\left( r^{2}+a^{2}\right) ^{2}-\Delta a^{2}\sin ^{2}\theta ,
\\
\Psi ^{2} &=&\Delta -a^{2}\sin ^{2}\theta , \\
\chi &=&2a\frac{GM}{c^{2}}r.
\end{eqnarray}

With the purpose of simplifying the calculations, we consider a circular and
equatorial orbit $r=1$ = constant, and $\theta =\frac{\pi }{2}$. Thus, using
equations $\left( \ref{4}\right) $ and $\left( \ref{5}\right) $ we find,
after a long calculation,

\vspace{1pt}

\begin{eqnarray}
&&c^{2}dt^{2}+2\,\frac{^{0}\Gamma _{03}^{r}}{^{0}\Gamma _{00}^{r}}%
cdtd\varphi +  \nonumber \\
&&+\frac{^{0}\Gamma _{33}^{r}}{^{0}\Gamma _{00}^{r}}d\varphi ^{2}\,-b\frac{1%
}{^{0}\Gamma _{00}^{r}}[g^{11}e^{(0)}\,_{0}T_{\left( 0\right)
01}dx^{0}dx^{0}+  \nonumber \\
&&+g^{11}e^{(1)}\,_{3}T_{\left( 1\right) 01}dx^{3}dx^{0}+  \nonumber \\
&&+g^{11}e^{(2)}\,_{3}T_{\left( 2\right) 01}dx^{3}dx^{0}+  \nonumber \\
&&+g^{11}e^{(1)}\,_{3}T_{\left( 1\right) 31}dx^{3}dx^{3}+  \nonumber \\
&&+g^{11}e^{(2)}\,_{3}T_{\left( 2\right) 31}dx^{3}dx^{3}+  \nonumber \\
&&+g^{11}e^{(3)}\,_{3}T_{\left( 3\right) 31}dx^{3}dx^{3}+  \nonumber \\
&&+g^{11}e^{(1)}\,_{0}T_{\left( 1\right) 01}dx^{0}dx^{0}+  \nonumber \\
&&+g^{11}e^{(2)}\,_{0}T_{\left( 2\right) 01}dx^{0}dx^{0}+  \nonumber \\
&&+g^{11}e^{(2)}\,_{0}T_{\left( 2\right) 31}dx^{0}dx^{3}+  \nonumber \\
&&+g^{11}e^{(1)}\,_{0}T_{\left( 1\right) 31}dx^{0}dx^{3}]  \nonumber \\
&=&0.  \label{6}
\end{eqnarray}
We have adopted Schwinger's time gauge \cite{8},

\begin{equation}
e^{(0)}\,_i =e_{\left( 0\right) i}=0,\;\;\;\; e^{\left( k\right) 0}=0 .
\end{equation}

In the case of asymptotically flat space-times, the tetrad fields that
satisfy Schwinger's time gauge condition, and the symmetric condition in
Cartesian coordinates $\cite{17a,17}$,

\begin{equation}
e_{(i)j}(t,x,y,z)=e_{(j)i}(t,x,y,z),  \label{7}
\end{equation}
are given by,

\begin{equation}
e_{a\mu }=\left( 
\begin{array}{cccc}
-\frac{1}{\rho }\sqrt{\Psi ^{2} +\frac{\chi ^{2}}{\Sigma ^{2}}\sin
^{2}\theta } & 0 & 0 & 0 \\ 
\frac{\chi }{\Sigma \rho }\sin \theta \sin \varphi & \frac{\rho }{\sqrt{%
\Delta }}\sin \theta \cos \varphi & \rho \cos \theta \cos \varphi & -\frac{%
\Sigma }{\rho }\sin \theta \sin \varphi \\ 
-\frac{\chi }{\Sigma \rho }\sin \theta \cos \varphi & \frac{\rho }{\sqrt{%
\Delta }}\sin \theta \sin \varphi & \rho \cos \theta \sin \varphi & \frac{%
\Sigma }{\rho }\sin \theta \cos \varphi \\ 
0 & \frac{\rho }{\sqrt{\Delta }}\cos \theta & -\rho \sin \theta & 0
\end{array}
\right) .
\end{equation}

Certainly there is a infinity of tetrads that yield the metric tensor $%
\left( \ref{8}\right) $, but only one that leads to the correct
gravitational energy description $\cite{17a}$.

Considering that

\begin{eqnarray}
^{0}\Gamma _{00}^{r} &=&-\frac{1}{2}g^{11}\partial _{r}g_{00}=\frac{GM}{%
c^{2}r^{4}}\left( r^{2}+a^{2}-2\frac{GM}{c^{2}}r\right) , \\
\frac{^{0}\Gamma _{03}^{r}}{^{0}\Gamma _{00}^{r}} &=&\frac{\partial
_{r}g_{03}}{\partial _{r}g_{00}}=-2a, \\
\frac{^{0}\Gamma _{33}^{r}}{^{0}\Gamma _{00}^{r}} &=&\frac{\partial
_{r}g_{33}}{\partial _{r}g_{00}}=\left( a^{2}-\frac{r^{3}c^{2}}{GM}\right) ,
\end{eqnarray}
and

\[
T_{\left( 0\right) 01}=\frac{1}{2}\left[ 1-\frac{2MG/c^{2}}{r}+\frac{%
4a^{2}G^{2}M^{2}/c^{4}}{\left[ \left( r^{2}+a^{2}\right) ^{2}-\left(
r^{2}+a^{2}-2GMr/c^{2}\right) a^{2}\right] }\right] ^{-1/2}\times 
\]

\begin{equation}
\times \left[ \left( \frac{2MG/c^{2}}{r^{2}}-\frac{4a^{2}G^{2}M^{2}/c^{4}[4%
\left( r^{2}+a^{2}\right) r-2\left( r-GM/c^{2}\right) a^{2}]}{\left[ \left(
r^{2}+a^{2}\right) ^{2}-\left( r^{2}+a^{2}-2GMr/c^{2}\right) a^{2}\right]
^{2}}\right) \right] ,
\end{equation}

\vspace{1pt}

\begin{equation}
T_{\left( 1\right) 01}=\left[ \frac{2a\frac{GM}{c^{2}}[4\left(
r^{2}+a^{2}\right) r-2\left( r-\frac{GM}{c^{2}}\right) a^{2}]}{\left(
r^{2}+a^{2}\right) ^{2}-\left( r^{2}+a^{2}-2\frac{GM}{c^{2}}r\right) a^{2}}%
\right] \sin \phi ,
\end{equation}

\vspace{1pt}

\begin{equation}
T_{\left( 2\right) 01}=\left[ \frac{2a\frac{GM}{c^{2}}[4\left(
r^{2}+a^{2}\right) r-2\left( r-\frac{GM}{c^{2}}\right) a^{2}]}{\left(
r^{2}+a^{2}\right) ^{2}-\left( r^{2}+a^{2}-2\frac{GM}{c^{2}}r\right) a^{2}}%
\right] \cos \phi ,
\end{equation}

\[
T_{\left( 1\right) 31}=-\frac{r\sin \varphi }{\left( r^{2}+a^{2}-2\frac{GM}{%
c^{2}}r\right) ^{1/2}}+ 
\]

\[
+\left\{ \frac{2\left( r^{2}+a^{2}\right) r-\left( r-\frac{GM}{c^{2}}\right)
a^{2}}{\left[ \left( r^{2}+a^{2}\right) ^{2}-\left(
r^{2}+a^{2}-2GMr/c^{2}\right) a^{2}\right] ^{1/2}}\right\} \frac{\sin
\varphi }{r}- 
\]

\begin{equation}
-\left\{ \frac{\left[ \left( r^{2}+a^{2}\right) ^{2}-\left(
r^{2}+a^{2}-2GMr/c^{2}\right) a^{2}\right] ^{1/2}}{r^{2}}\right\} \sin
\varphi ,
\end{equation}

\[
T_{\left( 2\right) 31}=\frac{r\cos \varphi }{\left( r^{2}+a^{2}-2\frac{GM}{%
c^{2}}r\right) ^{1/2}}- 
\]

\[
-\left\{ \frac{2\left( r^{2}+a^{2}\right) r-\left( r-\frac{GM}{c^{2}}\right)
a^{2}}{\left[ \left( r^{2}+a^{2}\right) ^{2}-\left(
r^{2}+a^{2}-2GMr/c^{2}\right) a^{2}\right] ^{1/2}}\right\} \frac{\cos
\varphi }{r}+ 
\]

\begin{equation}
+\left\{ \frac{\left[ \left( r^{2}+a^{2}\right) ^{2}-\left(
r^{2}+a^{2}-2GMr/c^{2}\right) a^{2}\right] ^{1/2}}{r^{2}}\right\} \cos
\varphi ,
\end{equation}

\begin{equation}
T_{\left( 3\right) 31}=0,
\end{equation}
we can rewrite equation $\left( \ref{6}\right) $, in the form

\begin{equation}
a^{\prime}c^{2}dt^{2}+b^{\prime}cdtd\varphi +c^{\prime}d\varphi ^{2}=0,
\label{10}
\end{equation}
where

\begin{equation}
a^{\prime}=1-b,
\end{equation}

\begin{equation}
b^{\prime }=-2a\left\{ 1-b\left[ 1+\frac{r^{2}}{\left[ \left( r^{2}+a^{2}-2%
\frac{GM}{c^{2}}r\right) \left( r^{2}+a^{2}+2\frac{GM}{c^{2}}\frac{a^{2}}{r}%
\right) \right] ^{1/2}}\right] \right\} ,
\end{equation}

\begin{equation}
c^{\prime }=\left\{ a^{2}-\frac{c^{2}}{\omega _{0}^{2}}\left[ 1+b\left( 1-%
\frac{GMa^{2}}{c^{2}r^{3}}-\left[ \frac{\left( r^{2}+a^{2}+2\frac{GM}{c^{2}}%
\frac{a^{2}}{r}\right) }{\left( r^{2}+a^{2}-2\frac{GM}{c^{2}}r\right) }%
\right] ^{1/2}\right) \right] \right\} ,
\end{equation}
and $\omega _{0}^{2}=\frac{GM}{r^{3}}$.

The square time interval $d\tau ^{2}$ is calculated by means of the line
element $\left( \ref{9}\right) $,

\begin{eqnarray}
\left( \frac{d\tau }{d\varphi }\right) ^{2} & =&\left( 1-\frac{2GM}{c^{2}r}%
\right) \left( \frac{dt}{d\varphi }\right) ^{2}+\frac{1}{c}\frac{dt}{%
d\varphi }\left( \frac{4GMa}{c^{2}r}\right) -  \nonumber \\
&&-\frac{1}{c^{2}}a^{2}\left( 1+\frac{2GM}{c^{2}r}\right) -\frac{r^{2}}{c^{2}%
}.  \label{11}
\end{eqnarray}
With the help of expression $\left( \ref{10}\right) $ we can write

\begin{equation}
c\frac{dt}{d\varphi }=\frac{-b^{\prime }\pm \left( {\ b^{\prime }}%
^{2}-4a^{\prime }c^{\prime }\right) ^{1/2}}{2a^{\prime }}.
\end{equation}
Substituting the equation above in expression $\left( \ref{11}\right) $ and
integrating in $\varphi $ from $0$ to $2\pi $, we find the square time
differences

\begin{equation}
\frac{\tau _{+}^{2}-\tau _{-}^{2}}{2T_{0}}\sim \frac{4\pi a}{c}\frac{\left(
1-2b\right) }{\left( 1-b\right) ^{3/2}},  \label{2003}
\end{equation}
in the limit $r\gg 2GM/c^{2}$ and $r\gg a.$ $T_{0}=\frac{2\pi }{\omega
_{0}\left( 1-b\right) ^{1/2}}$ is the orbital period. For $b=0$ the
expression ($\ref{2003}$) coincides with the one found by Mashhoon {\it et al%
}. $\cite{1},$ who considered clocks as point particles, i.e., without
internal structure.

For $b=1$ the first term containing $(1-b)c^{2}dt^{2}$ in equation $\left( 
\ref{10}\right) $ vanishes, a fact that prevents from recovering the general
relativity limit. By keeping $b\neq 1$ in the limit $r\gg 2GM/c^{2}$ and $%
r\gg a$, we obtain the periods for a direct and retrograde orbit

\begin{equation}
\tau _{\pm }\sim \frac{2\pi }{\omega _{0}\left( 1- b\right) ^{1/2}}\pm \frac{%
2\pi a}{c}\frac{\left( 1-2b\right) }{\left( 1- b\right) }.  \label{13}
\end{equation}
For $b=0$ we obtain the usual result of general relativity,

\begin{equation}
\tau _{\pm }\sim \frac{2\pi }{\omega _{0}}\pm \frac{2\pi a}{c}.
\end{equation}

From equation $\left( \ref{13}\right) $ we conclude that

\begin{equation}
\tau _{+}-\tau _{-}\sim \frac{4\pi a}{c}\frac{\left( 1-2b\right) }{\left(
1-b\right) }.  \label{2005}
\end{equation}

The presence of the factor $\left( 1-b\right) ^{1/2}$ in the first term of
expression $\left( \ref{13}\right) $ suggests that the Newtonian
gravitational potential is modified according to

\begin{equation}
\phi (r)=-\frac{GM}{r}(1-b).  \label{15}
\end{equation}
This result agrees with that obtained by Wanas in a completely different way 
$\cite{18}$. Mashhoon {\it et al}. $\cite{1}$ considered point like clocks,
without internal structure. Such clocks cannot couple with the space-time
torsion. An atomic clock certainly has internal structure, and therefore
spin.

\bigskip

\section{Estimative of the empirical parameter}

\bigskip An estimative of the parameter $b$ can be made by taking into
account Einstein's equivalence principle. According to the latter, (a) the
trajectory of a freely falling body is independent of its internal structure
and composition (known as weak equivalence principle), and (b) the outcome
of any local non-gravitational experiment is independent of the velocity of
the freely-falling reference frame in which it is performed, and of its
position in space and time (local position invariance).

The gravitational redshift of spectral lines is ultimately due to Einstein's
equivalence. This effect is universal and independent of the nature of the
clock, and is given by

\begin{equation}
\nu =\nu _{0}\left( 1+\frac{\phi _{N}}{c^{2}}\right) ,
\end{equation}
where $\nu _{0}$ is the proper clock frequency when the $\phi _{N}=0$ and $%
\nu $ is the frequency redshifted by the gravitational potential $\phi _{N}.$
In the present work the Newtonian gravitational potential is modified by the
parameter $b$. With the help of equation ($\ref{15}$) we can write the
corresponding result in the context of our analysis,

\bigskip

\begin{equation}
\nu =\nu _{0}\left[ 1+\left( 1-b\right) \frac{\phi _{N}}{c^{2}}\right] .
\end{equation}
Therefore we may determine the parameter $b$ by looking for experiments that
violate the equivalence principle.

In the last decades, possible violations of the equivalence principle were
tested by means of experiments related to the violation of local position
invariance. When the local position invariance principle\ is violated, the
frequency is expected to be $\cite{4}$

\begin{equation}
\nu =\nu _{0}\left[ 1+\left( 1+\beta \right) \frac{\phi _{N}}{c^{2}}\right] ,
\end{equation}
$\beta $ being a dimensionless parameter (positive or negative) that
presents a dependence on the internal structure of the clock, and that
measures the local position invariance violation of the clock in
consideration. Therefore the determination of the parameter $b$ amounts to
the the fixation of the parameter $\beta $.

Experiments with two clocks have been carried out with the purpose of
measuring the difference between the two frequencies,

\begin{equation}
\left( \nu _{2}-\nu _{1}\right) /\nu _{0}=\left( 1+\beta \right) \left[ 
\frac{\left( \phi _{N2}-\phi _{N1}\right) }{c^{2}}\right] ,
\end{equation}
where two identical clocks, $1$ and $2$, experience different gravitational
potentials, $\phi _{N1},$ and $\phi _{N2}$, respectively. In the Table, we
display some of the results of the experiments that determine the parameter $%
\beta $. The Table shows experiments performed with two Cesium atomic clock
and two H masers.

\bigskip

\begin{center}
Table. Some experiments with two identical clocks to determination of the $%
\beta $ parameter $\cite{19}$

\begin{tabular}{lll}
Clocks & $\beta $ & Reference \\ 
H-H & $\left| \beta _{H}\right| <7\times 10^{-5}$ & Vessot {\it et al.} $
\cite{20}$ \\ 
Cs-Cs & $\left| \beta _{Cs}\right| <10^{-1}$ & Hafele and Keating $\cite{21}$
\\ 
Cs-Cs & $\left| \beta _{Cs}\right| <1.5\times 10^{-2}$ & Alley $\cite{22}$
\\ 
Cs-Cs & $\left| \beta _{Cs}\right| <2\times 10^{-1}$ & Briatore and
Leschiutta $\cite{23}$ \\ 
Cs-Cs & $\left| \beta _{Cs}\right| <6\times 10^{-2}$ & Iijima and Fujiwara $
\cite{24}$%
\end{tabular}
\end{center}

The determination of the parameter $\beta _{H}$ resulted of the test of
redshift based on the measurement of the frequency shift of a H maser on a
spacecraft launched upward to 10.000 km compared with a similar maser on
Earth. This is one of the most precise experiments about redshift performed
so far. The internal structure of the H maser is simpler than that of the
Cesium atomic clock. Taking the values of the parameter $\left| \beta
_{H}\right| $ of reference $\cite{20}$ as values of our parameter $b$, we
can calculate the time difference using equation $\left( \ref{2005}\right) $%
. It turns out that there is a difference of 0.007\% with respect to
Mashhoon's result. The use of $\left| \beta _{Cs}\right| $ of reference $
\cite{22}$ results in a difference of 1.52\% with respect to the expected
value.

The parameter $b$ can also be estimated by the redshift experiment conducted
by Pound and Snider $\cite{25}$ that measured the frequency shift of
gamma-ray photons from $^{57}$Fe as the result of M\"{o}ssbauer effect. In
our understanding, the internal structure of $^{57}$Fe can couple with the
torsion field. The measurement of the redshift yield the value ($0.9990\pm
0.0076)\times 4.905\times 10^{-15}$ predicted by the equivalence principle.
Our prediction is

\begin{equation}
(1-b)\times 4.905\times 10^{-15},
\end{equation}
and therefore $b$ $<(0.0010\pm 0.0076)\approx 7.6\times 10^{-3}.$This result
is in agreement with the results of the Table displayed above for other
types of atoms, and in particular with those of Ref. $\cite{22}$. Such
value of $b$ yields a difference of 0.77\% with respect to the general
relativity prediction.

The equivalence principle can be tested using two nonidentical clocks in the
same gravitational potential $\cite{19}$, e.g., the Cs clock and Mg clock.
In this experiment, it is measured the difference between the parameters $%
\left| \beta _{Cs}-\beta _{Mg}\right| <7\times 10^{-4}$ that represent the
coupling of the hyperfine and fine-structure transition in these atoms. This
result may indicate the dependence of the fine-structure constant with the
gravitational potential. A further experiment with two nonidentical clocks
yield $\left| \beta _{Cs}-\beta _{H}\right| <2.1\times 10^{-5} \cite{26}$.

\section{Weak field approximation}

Making use of a Lagrangian quadratic in the torsion tensor, constructed in
terms of three free parameters $c_{1},$ $c_{2}$, and $c_{3}$, and
considering the weak field approach, Hayashi and Shirafuji $\cite{9}$ wrote
down the geodesic equation for a particle in the weak field approximation
according to

\begin{equation}
\frac{d^{2}x^{i}}{dt^{2}}=\frac{2}{9}k\frac{\left( c_{1}+4c_{2}\right) }{%
c_{1}c_{2}}\frac{\partial }{\partial x^{i}}\phi _{N},\;\;\;\;i=1,2,3,
\end{equation}
where$\ k={c^{3}}/(16\pi G)$. A particular combination of the parameters $%
c_{1},$ $c_{2}$, and $c_{3}$ leads to the condition for the Newtonian limit,

\begin{equation}
c_{2}=-\frac{\left( c_{1}+\frac{2}{3}k\right) }{\left( 1+\frac{9}{8k}%
c_{1}\right) }+\frac{2}{3}k.
\end{equation}

Wanas {\it et al.} $\cite{6}$ proposed the particle equation in the weak
field limit of the gravitational field to be

\begin{equation}
\frac{d^{2}x^{i}}{dt^{2}}=-\frac{\partial }{\partial x^{i}}\phi _{s},
\;\;\;\;\;i=1,2,3,
\end{equation}

\begin{equation}
\phi _{s}=\left( 1-b\right) \phi _{N},
\end{equation}
where $b$ acquires the values

\begin{equation}
b=\frac{n}{2}\alpha \gamma ,\;\;\;\;\;\;n=0,1,2,3...
\end{equation}
and $\alpha =1/137$ is the fine structure constant; $\gamma $ is an
adjustable parameter to be fixed by the experience. According to
interpretation of Wanas {\it et al.}, depending on the value of $n$ we have
particles with spin $0,1,2,3$ and so on.

We suggest that the interpretation of Hayashi and Shirafuji, and of Wanas 
{\it et al.} may be reconciled by writing

\begin{equation}
b=1+\frac{2}{9}k\frac{\left( c_{1}+ 4c_{2}\right) }{c_{1}c_{2}}.  \label{17}
\end{equation}

We already know that in order to have a well defined Hamiltonian formulation
(in the time gauge condition) it is necessary to have two extra conditions
on the parameters $c_{1},$ $c_{2}$ and $c_{3}$ $\cite{10}$,

\begin{equation}
c_{1}+c_{2}=0,  \label{16}
\end{equation}
and

\begin{equation}
c_{1}=\frac{4}{9}c_{3}.  \label{18}
\end{equation}
From equations $\left( \ref{17}\right) $, $\left( \ref{16}\right) $ and $%
\left( \ref{18}\right) \ $ we obtain

\begin{eqnarray*}
c_{1} &=&-\frac{2}{3}k\frac{1}{1-b}, \\
c_{2} &=&\frac{2}{3}k\frac{1}{1-b}, \\
c_{3} &=&-\frac{3}{2}k\frac{1}{1-b}.
\end{eqnarray*}
For $b=0$ (i.e., particles without intrinsic spin) we find that the
parameters lead to the teleparallel equivalent of general relativity $\cite
{10},$ $\cite{13}$. A nonzero value of $b$ establishes a connection between
the TEGR and the geodesic equation that will eventually match with the
experiments. Note, however, that according to Hayashi and Shirafuji the
experiments do not confirm that $c_{i}$ have the exact values given above
with $b=0$.

\section{Conclusions}

In this work we suggest that there is a connection between the following
different issues:

a) the fixation of a global Lorentz symmetry,

b) the gravitomagnetic effect,

c) the absolute parallelism of Einstein,

d) the equivalence principle,

e) the intrinsic spin of the particles (a quantum aspect) and the space-time
torsion (a classical aspect),

f) the teleparallel equivalent of general relativity and the conditions of
Legendre transform that guarantee a well defined time evolution in the
Hamiltonian framework, and

g) Schwinger's time gauge condition.

We concluded that it is possible to have a modified geodesic equation, and
investigated it in the context of the gravitomagnetic effect. We also
concluded that is possible to describe (on phenomenological grounds) the
spin-torsion interaction, by introducing a small correction to the geodesic
equation. For macroscopic bodies and particles without spin, this effect
does not occur. The small value of $b$ does not invalidate Einstein's
general relativity, because the geodesic equation does not depend on
Einstein's equations. All results of the teleparallel equivalent of general
relativity remain valid in the limit $b=0.$ The existence of a small
correction suggests a small violation of the principle of equivalence that
could be determined experimentally.  Future space experiments
will indicate the correct value of $b$. The violation of the
local position invariance measured by parameter $\beta $ could
be explained by the interaction of the
internal composition of the clocks with the torsion field.

It is possible that there exists a relation between the empirical coupling
constant $b$ and the contortion tensor in equation $\left( \ref{3}\right) .$
The relation between the contortion tensor with the intrinsic spin of the
particles (electrons) is suggested in Hayashi and Shirafuji's
work $\cite{9}$.

It remains to discover the variational principle that leads to the correct
geodesic equation. Also, it is necessary a further understanding of the
coupling between the contortion tensor and the spin of matter. Efforts in
this respect will be carried out.

\bigskip \noindent {\it Acknowledgements}

\noindent One of us, A. A. Sousa, is grateful to Faculdades Planalto for
financial support.


\end{document}